\documentstyle[preprint,prl,aps]{revtex}

\begin{document}
\draft
\preprint{}
\title{Electronic excitation spectra from {\it ab-initio} band
structure results for La$M$O$_3$ ($M$ = Cr-Ni)}
\author{D. D. Sarma, N. Shanthi and Priya Mahadevan}
\address{Solid State and Structural Chemistry Unit, Indian Institute of 
Science, Bangalore 560012, India}
\date{\today}
\maketitle
\begin{abstract}

We present calculated electron excitation
spectra for the La$M$O$_3$ series ($M$=Cr-Ni) 
obtained within {\it ab-initio} band structure calculations for the
real geometric and magnetic structures. The calculated results 
show good agreement with the experimentally obtained
spectra. This suggests that the transition metal-transition
metal interactions via the oxygen atom play an important
role in determining the spectroscopic features 
and indicates a smaller
value of $U/t$ than has been believed so far. The present
approach undermines the importance of multiplet interactions
which otherwise play an important role within various 
single impurity models.

\end{abstract}
%\narrowtext
%\twocolumn

\noindent {\bf Introduction}

During the last few years, a lot of attention has been focussed on
the electronic structure of the 3$d$ transition metal oxides
following the discovery of many exotic properties like
high-temperature superconductivity \cite{BM}, giant magnetoresistance
\cite{GMR} and other phenomena \cite{Tokura}. Many of these
properties indicate a close interplay between magnetic and electronic
properties. This arises from a simultaneous presence of strong
electron-electron interaction strength, $U$, within the
transition metal $d$ manifold, and a sizable hopping
interaction strength, $t$ between the transition metal
$d$ and oxygen $p$ states. While large values of $U$ tend to localize
the electrons and stabilize a magnetic moment at the transition metal
site, the presence of large $t$, on the other hand, leads to a
tendency towards delocalization of the electrons. Thus, this
competition between the two opposing tendencies leads to the wide
spectrum of electronic and magnetic ground states in transition
metal oxides. Electron spectroscopies have played a very important
role in understanding the electronic structure of such strongly
correlated electron systems over more than a decade now. These
spectroscopic results have been traditionally interpreted with the
help of various single-impurity models, such as the cluster model
\cite{Cluster} or the Anderson impurity model \cite{Imp}. Within this
approach, one considers only a single transition metal ion
interacting with a cluster of oxygen atoms \cite{Cluster}, or with a
ligand derived $p$-band \cite{Imp}. Thus, this model completely
ignores the lattice periodicity of the transition metal site. It is
reasonable to expect a negligible strength of direct hopping
interaction between nearest neighbor transition metal sites owing to
their large spatial separation in most of the oxides; however, two
transition metal sites can still have a substantial effective interaction
 via the
oxygen atom in between, if the transition metal $d$ - oxygen $p$
hopping interaction strength, $t$ is large enough. Thus, it is
obvious that a single impurity approach is expected to be most
suitable in the limit of large $U/t$. While it is generally assumed
to be the case, it is, however, not yet an
established fact for {\it all} transition metal oxides.

In very recent times, there is a growing awareness that there may indeed
be a strong coupling between the neighboring transition metal sites
in many such compounds, making it necessary to explore models that
are more relevant to the electronic structure of transition metal
oxides than a single-impurity model. For example, it has been
recently shown that the spectral features of the prototypical
examples of strongly correlated transition metal oxides, NiO and CuO
\cite{Multi} are better described within a multi-impurity model
compared to single-impurity ones. More recently, it has been shown
\cite{PRL} that the ground state magnetic and electronic properties
of La$M$O$_3$ compounds with $M$ = Mn-Ni are correctly predicted by
{\it ab-initio} band structure calculations. 
Furthermore, it was found that the {\it ab-initio} 
band structure approach also provided a
reliable description of the x-ray photoelectron
spectra of the valence band in La$M$O$_3$ with $M$=Mn-Ni \cite{PRL}.
In contrast, the traditional belief has been that band theory
is a complete failure for the excitation spectra of every 
kind of transition metal compound and that it is 
necessary to describe the excitation spectra within models
appropriate for strongly correlated systems, such as the
Anderson impurity model. It is to be noted that the band theory
and the Anderson impurity model represent two attempts to describe
correlated systems from two different limits. While the band
theory approach provides an accurate and {\it ab-initio}
description of various hopping interaction strengths
in the lattice, it treats the Coulomb interaction
only in an approximate and average effectively 
single particle sense. On the other hand, impurity
hamiltonians ignore the lattice of the transition metal
ions, while providing a rigorous though parametrized treatment
of the intra-atomic Coulomb interaction. It is well-known
that the impurity model approach has been satisfactory
in general in describing a variety of excitation spectra
of a large number of transition metal compounds. On the 
other hand, the success of band theory to describe the
excitation spectra of La$M$O$_3$ compounds with $M$=Mn-Ni
\cite{PRL} has been demonstrated only for the x-ray valence 
band photoemission spectra. It is to be noted that such spectra
are related only to the occupied parts of the electronic 
structure. In view of these considerations,
the unexpected success of the band theory in this context
is worth a closer inspection, since there are certain
obvious merits in an {\it ab-initio} approach
compared to any parametrized model. Thus, it was felt
necessary to investigate whether band theoretical
approach would also describe the unoccupied parts of
the electronic structure in the same compounds, thus
providing a unified and {\it ab-initio} description
of excitation spectra related to 
the entire electronic  structure of these compounds. In 
order to do this, here we report the calculated unoccupied
spectra for $M$=Cr-Ni and compare these with corresponding
Bremsstrahlung isochromat (BI) spectra which probe the
unoccupied part. In a few cases, the experimental BI
spectra do not provide a critical check for the
calculated spectra due to experimental difficulties. 
In such cases, we compare the oxygen $K$-edge 
x-ray absorption (XA) spectra with the corresponding calculated
ones, since XA spectra also probe the unoccupied part of the electronic
states. 
Furthermore, we present new results for the x-ray
photoemission (XP) spectrum of LaCrO$_3$, not presented
earlier \cite{PRL}. All these results 
are discussed in terms of spin-polarizations and orbital
characters of the states. 

\noindent {\bf Method}

{\it Ab-initio} LSDA calculations were carried out using
linearized muffin-tin orbital (LMTO) method 
within the atomic sphere approximation (ASA) \cite{LMTO}.
For the series of compounds investigated, no empty sphere
was required to fulfil the volume-filling criterion.
Semi-relativistic calculations with full self-consistency
were obtained using $s,p,d$ and $f$ basis at each atomic sphere.
The calculations were performed with 283 and 80 k-points
in the irreducible parts of the Brillouin zones (BZ) of the
rhombohedral and orthorhombic structures, corresponding to 1000 and 216
k-points over the full BZ. In order to obtain the calculated
spectra, it is necessary to calculate the transition
matrix elements, since the spectral features are substantially
modified from the DOS due to the presence of strong
variations in transition probabilities corresponding to states
with different site and angular momentum characters as well as
with energy. Thus, energy dependent matrix elements were
calculated for each angular momentum $l$ and site starting with
the converged LMTO potentials for each site within the
formalism of Winter {\it et~al.} \cite{Winter}. 
In order to obtain the calculated
spectra, the site and $l$-projected DOS were weighted by these calculated
matrix elements and suitably broadened with a Gaussian 
representing the resolution broadening and an energy dependent
Lorentzian representing life-time effects.
Since the oxygen $K$-edge x-ray absorption spectrum probes the
oxygen $p$-character in the unoccupied DOS due to dipole
selection rules, only the unoccupied oxygen $p$ partial
DOS without any matrix element correction was taken into
account in this case. The calculated spectra were rigidly
shifted in energy to align the most intense peak with that 
of the experiment correcting for the underestimation of the band gap
in local density approximation, as has been discussed earlier
\cite{PRL}.

Every calculation was performed for the real crystal structure
\cite{structures} and the observed magnetic 
ground state. LaCrO$_3$ and LaFeO$_3$ have twenty 
atoms in the Pbnm unit cells; LaMnO$_3$ also has twenty atoms in the
Pnma unit cell.  
LaCoO$_3$ and LaNiO$_3$ have ten atoms each in the R$\bar{3}$c structure.
The magnetic 
structure for which calculations were performed for these
oxides are $A$-type AF (Mn), $G$-type AF (Cr and Fe)
and nonmagnetic (Co and Ni) according to the observed
ground state  magnetic structures.

\noindent {\bf Results and discussion}

We show the comparison of x-ray photoemission
valence band spectrum of LaCrO$_3$ and the corresponding
calculated spectrum in Fig. 1. The experimental spectrum
exhibits the most intense peak at about 6 eV with a
shoulder at about 4.5 eV binding energy. Another peak
is clearly observed at about 1.5 eV binding energy. The spectral
intensity of this feature relative to the most intense
peak at 6 eV is found to become weaker with decreasing
photon energy. This indicates that the spectral
feature at 1.5 eV is predominantly due to Cr states, while
the main peak has more oxygen $p$ character. The
calculated spectrum shown in Fig. 1a is in very  good 
agreement with the spectral
feature at about 6 eV and the shoulder at about 4.5 eV both in terms
of spectral features, widths, energy positions and relative
intensities. However, the energy position of the Cr $d$ related
feature near 1.5 eV is underestimated by about 0.9 eV in the calculation,
though the spectral shape, intensity and width appear to be reasonably
well described. The Cr $d$ and O $p$ contributions to the calculated 
spectrum shown in Fig. 1a show that the experimental feature at about
1.5 eV is almost entirely contributed by Cr $d$ states, while the
oxygen $p$ related intensity is the dominant contribution to the
most intense peak and the shoulder at higher binding energy,
in agreement with the dependence of the observed
experimental intensity 
on photon energy. The calculated DOS
and partial DOS suggest that the shoulder at about 4.5 eV
arises primarily from non-bonding oxygen $p$ states with minimal contribution
from the Cr $d$ states, while  
the most intense peak
arises from bonding Cr $d$-O $p$ interaction.
This interpretation is supported by the distribution of Cr $d$ intensity
in the calculated spectrum in Fig. 1a between 3 and 8 eV. 
The spin polarization of Cr $d$ contribution from one of the Cr-sites
in the calculated spectrum shows a very strong polarization
for the primarily Cr $d$ related intensity between 0 and 2 eV; however, 
the extent of spin polarization is considerably
weakened in the energy region of primarily oxygen $p$
contribution between 3 and 8 eV. The strong spin-polarization
of the Cr $d$ related feature is easily understood in terms of the 
$t_{2g \uparrow}^3$ configuration of Cr$^{3+}$ $d^3$ state.

It was not possible to record the BI spectrum of LaCrO$_3$ due to
severe charging problem arising from the wide band gap insulating
state of the compound. Doping of 10\% Sr in LaCrO$_3$ decreases the
resistivity of the sample considerably. The spectral features
of La$_{0.9}$Sr$_{0.1}$CrO$_3$ have been found to be very
similar to that of LaCrO$_3$ in various photoemission
spectra and we believe that the BI spectra
of these two compounds are also similar. We show the wide
scan BI spectra of La$_{0.9}$Sr$_{0.1}$CrO$_3$ in Fig. 2. The intense
peak near 8 eV arises from La 4$f$ contribution, while the strongly
distorted conduction
band spectral features can be observed only as a shoulder
between 2 and 6 eV. This illustrates the problem of
extracting any detailed spectral feature of the conduction
band in such compounds due to the overlap of very intense
features in BI spectra at higher energies. We do not believe that any
spectral decomposition in such a situation will be reliable in
absence of any detailed knowledge of the spectral shape of other
features and thus, we do not make any such attempt here. Instead we
compare the calculated spectrum of the unoccupied state with the raw
experimental results in Fig. 3a.  The calculated spectrum exhibits a
peak close to 4 eV in the Cr $d$ derived states.  The calculated
feature agrees with the experimental spectrum as suggested in Fig. 3a
by a vertical line. However, the superposed features arising from
other states, for example La 4$f$ states, make the determination of
the experimental peak position very uncertain.  The spectral
decomposition in terms of the Cr $d$ and O $p$ contributions (Fig.
3a) indicate that there is  very little contribution from O
$p$-derived states in the BI spectrum.  The calculated peak at about
4 eV has more down-spin character, though the up-spin contribution is
also sizable (see Fig. 3b). The relative intensities of the
spin-resolved contributions suggest the down-spin to be arising from
the transitions into the $t_{2g\downarrow}^0$ states, while the
up-spin contribution in the same energy region arises from
$e_{g\uparrow}^0$ states.  There is a second spectral feature related
to Cr $d$ states in the BI spectrum; this can be barely seen as an
asymmetry in the low energy wing of the La 4$f$ signal at about 6 eV
in Fig. 2. Our calculations show that it has substantial Cr $d$
down-spin contributions, though other states such as La $d$
also significantly  contribute
broad spectral features at this energy range.
We interpret the Cr $d$ contribution near 6 eV as arising from 
transitions into $e^0_{g\downarrow}$ states in view of the spin-
polarization of Cr $d$ states in this energy range.
The energy separation between the  $t_{2g\downarrow}$
and $e_{g\downarrow}$
states is about 2.5 eV and is an approximate measure of the total 
crystal field splitting in LaCrO$_3$.

It is clear from Figs. 2 and 3 that the band theoretical
approach to describe the unoccupied spectral features
is severely restricted in the case of BI spectrum of 
LaCrO$_3$ due to the presence of the intense
La 4$f$ spectral contributions. In order to be more
certain of the efficacy of this approach, we
have compared the oxygen $K$-edge x-ray absorption spectrum
of La$_{0.9}$Sr$_{0.1}$CrO$_3$ with the 
broadened oxygen $p$-DOS of LaCrO$_3$ over a wider energy
range, since
the oxygen $K$-edge absorption spectra probe the oxygen
$p$-admixture in the unoccupied parts due to dipolar
selection rules. The calculated results suggest
that all the energy positions of various features
in the calculation are systematically overestimated by about 
25\% compared to the experimentally observed ones.  Such 
an overestimation is not unexpected in a linearized band structure 
approach, since the band structure equations are 
linearized near the mean energy of
the {\it occupied} densities of states in order to obtain accurate
descriptions of the potentials and various ground state properties.
Similar effects have been observed while comparing
the unoccupied states with experimentally obtained spectra of other
compounds as well \cite{estimate}. 
Thus, we have uniformly contracted the energy
scale of the
calculated results by 25\% before comparing it to the
experimentally obtained spectra in Fig. 4. In the same figure,
we also show the spin-polarized Cr $d$ partial DOS
after broadening for comparison.
The experimental
spectrum clearly shows a peak at about 4 eV due to 
transitions arising from
 $t_{2g\downarrow}$
and $e_{g\uparrow}$ states as evidenced from the spin-polarized
Cr $d$ state contributions. This is in good agreement
with the BI spectra presented in Fig. 3. The spectral feature
due to the e$_{g\downarrow}$ state is seen at about 6.5 eV.
The clearer emergence of these peaks in the 
oxygen $K$-edge spectrum compared to the BI spectrum
in Fig. 2 is due to the near absence of La
4$f$ related states in the $K$-edge spectrum; La
4$f$ states have relatively small oxygen $p$-admixture in these
compounds, while it dominates the BI spectrum. 
The calculated spectrum for
oxygen $K$-edge absorption in Fig. 4 clearly shows the first peak
arising from an oxygen $p$-admixture into the Cr $t_{2g\downarrow}$
and $e_{g\uparrow}$ states at the right energy position
with correct spectral shape and width.  The second feature in the
calculated spectrum arises from the oxygen $p$ admixture into 
the Cr e$_{g\downarrow}$ states.
The third spectral feature in the experiment and the 
calculation at about 8.5 eV arises mainly from oxygen $p$-admixture
with La derived states with negligible Cr $d$ contribution.
While there is a good overall agreement between the
experimental and calculated results, there is a discrepancy 
in terms of the relative 
intensities of the two peaks at about 6.5 and 8.5 eV. This arises
from our neglect of matrix element effects in the oxygen $K$-edge
absorption spectrum. It is to be noted that the feature
at 6.5 eV arises from oxygen $p$-admixture with Cr $d$ $e_{g\downarrow}$
states, while the higher energy feature at 8.5 eV is due to the
admixture with La states; 
thus the matrix element effects may be expected to play some role.
Overall it appears that the band structure results provide a
satisfactory and {\it ab-initio} description of the x-ray
absorption spectrum of this compound.

The BI spectrum of LaMnO$_3$ is compared with the calculated spectrum
in Fig. 5a. From this figure, it is obvious that the spectral shape,
width and position are well described by the calculated result. The
increasing intensity of the experimental spectrum above 3.5 eV arises
from the tailing of the intense La 4$f$ state, as has already been
discussed. We find that the BI spectrum is almost entirely dominated
by Mn $d$ derived states; this is a consequence of much larger matrix
elements for these states compared to oxygen $p$-states. The
spin-polarized contributions shown in Fig. 5b 
suggest a dominant contribution from
down-spin states arising from transitions into $t_{2g\downarrow}^0$ and
$e_{g\downarrow}^0$ states, though there is considerable up-spin
contribution in the first few electron-volts from transitions into
e$_{g\uparrow}^1$ state.

In the case of BI spectrum from LaFeO$_3$, we once again encounter
the same difficulty as in the case of LaCrO$_3$,
arising from the large band-gap and very insulating nature of this
compound. The consequent charging effects lead to 
a broadening of various spectral features
and an overlap of the La 4$f$ states from the higher energy
side, strongly distorting the Fe $d$-related spectral features.
However, the spectrum shown in Fig. 6a suggests 
the existence of two features,
marked by vertical lines. The positions of these features
are consistent with the calculated doublet features which 
are dominated by Fe $d$ states due to matrix element effects. The first
peak in the calculated spectrum corresponds to transitions
into $t_{2g\downarrow}$ state while the higher energy,
weaker intensity feature corresponds to the $e_{g\downarrow}$
state.
The spin-polarized contributions in Fig. 6b suggest a
nearly complete spin-polarization of the BI spectrum from
each Fe sites due to the  
$t_{2g\downarrow}^{3}$ $e_{g\downarrow}^{2}$
configuration of Fe$^{3+}$ high-spin state. In 
order to further test the usefulness of the band
structure approach in this case, we have also calculated
the oxygen $K$-edge x-ray absorption spectrum, in the same way as it was
done for LaCrO$_3$.  We show the experimental and calculated
spectra in Fig. 7, after contracting the energy scale of
the calculated result by 25\%, as before. 
The calculated spectrum is evidently in good agreement with the experimentally
observed features. The spin-polarization of the Fe $d$ states
in the same energy interval shown in Fig. 7 shows a doublet
feature with almost entirely down-spin character. The relative 
intensities of these two features clearly suggest that  
the low energy peak arises from the transitions
into $t_{2g}$ states, while the higher energy feature
arises from $e_g$ states, suggesting a crystal-field
splitting of about 1.4 eV in LaFeO$_3$. It is interesting
to note that while the intensity ratio of the two Fe
$d$ related features follow the expected degeneracy
ratio between the $t_{2g}$ and $e_{g}$ states, the
experimental as well as the calculated x-ray absorption spectra
have nearly equal intensities in the two 
corresponding features. This arises from the
fact that the oxygen  $p$ admixture in the
primarily Fe $d$ derived states, probed by 
$K$-edge XA spectroscopy, is substantially
larger for the $e_{g}$ symmetry arising
from a correspondingly larger hopping
interaction strength than that for the $t_{2g}$ symmetry \cite{Fe}.

We compare the experimental
and calculated BI spectra of LaCoO$_3$ in Fig. 8. The experimental spectrum
clearly shows a doublet feature,
though the spectral features are strongly distorted by the overlap
of signals from other states, such as La 4$f$. The calculated 
result describes the doublet structure well in terms of
energy positions and widths. The BI spectrum is found to be
dominated by Co $d$-derived states. 
Since the band structure calculation was 
performed for the nonmagnetic ground
state of LaCoO$_3$, there is no spin-polarization of the Co $d$
band in this case. It is to be noted that the Co $d$
 spectral feature in the BI spectrum arises from the
 transition   $t_{2g}^{6}$ $e_{g}^{0}$ to $t_{2g}^{6}$ $e_{g}^{1}$
and can give rise to only a single line within a single-impurity
approach corresponding to the transition from $^{1}A_{1g}$ to 
$^{2}E_{g}$ and the spectral width will be resolution
limited. In contrast, the spectral feature clearly suggests a two-peak
structure with considerable energy spread. 
Within the present
interpretation, the doublet feature and the spectral
width reflect features in the density of states 
obtained from band structure calculations \cite{Springer}. Likewise,
the x-ray photoemission spectrum of LaCoO$_3$ \cite{Laco}
shows a single narrow peak feature at 1 eV arising
 from the low-spin $t_{2g}^{6}$ configuration
of Co$^{3+}$ ions in an octahedral site. The width of the signal,
apart from the resolution broadening, is due
to the bandwidth of the $t_{2g}$ states arising from the 
Co-O-Co interactions. Within a single-impurity
description \cite{Abbate}, a photoemission transition 
from the $^{1}A_{1g}$ ($t_{2g}^{6}$) configuration
 can give rise to only a single line corresponding to a
 final state $^{2}T_{2g}$ ($t^{5}_{2g}$).
 Thus, the width of the spectroscopically observed
 feature will be limited only by the resolution 
 function of the spectrometer. However, the width
 experimentally
 observed appears to be considerably larger,
 suggesting limitations of a single-impurity description
 also for the photoemission spectrum.

Fig. 9 shows the experimental
and calculated BI spectra of LaNiO$_3$. The experimental spectrum
exhibits a rather broad single peak feature which appears to be
fairly well described by the calculation. However, the calculated
spectral width appears to be somewhat larger than the experimentally
observed one. One contributing factor to this mismatch is the 
systematic overestimation of unoccupied energies within the
linearized method, as has already been discussed. However, we note that
the existence of strong correlation effects within the metallic
conduction band derived from the interaction between Ni $d$ and O $p$ 
states, may also lead to a narrowing of the experimental feature.

The above results conclusively establish that the excitation
spectra of La$M$O$_3$ compounds with $M$=Cr-Ni are well described
by band calculations. In contrast, it is well known that band
calculations do not provide the correct description
for the excitation spectra of the corresponding $M$O
compounds. Thus, the present results suggest that the $U/t$
values are smaller in the La$M$O$_3$ compounds
compared to the corresponding $M$O compounds. 
This suggestion is indeed supported by several analysis
already available in the literature
\cite{Saitoh,Bocquet,Mizo,Sarma and others}.
For example, $U/t$ for NiO has been estimated to be 5.8
\cite{Saitoh} and 5.3 \cite{Szy} while the same for LaNiO$_3$ 
has been estimated to be about
3.9 \cite{Mizo} and 2.5 \cite{Sarma and others}.
Similarly, the $U/t$ for MnO is about 6.4 \cite{Saitoh}
whereas the same for LaMnO$_3$ is estimated to be 4.2
\cite{Saitoh} and 1.8 \cite{Sarma and others}; for TiO
it is 3.8 \cite{Saitoh} and for LaTiO$_3$ 1.7 \cite{Saitoh}.
Thus, we find that the $U/t$ for the $M$O compounds is always
significantly higher (about 50-100\%) than those for the La$M$O$_3$
compounds. This is probably the single most important reason
why band theory provides a satisfactory  description for the
electronic structure of La$M$O$_3$, while failing 
in the case of $M$O compounds.

In summary, we have compared various single-particle excitation 
spectra of La$M$O$_3$ compounds with $M$ = Cr-Ni to
those obtained from {\it ab-initio} band structure results. 
The calculated spectral features agree remarkably well with the experimental
ones for the wide range of compounds investigated. 
The present results clearly establish the
importance of a sizable hopping interaction between
various transition metal sites mediated via the oxygen
states.  
One of the main drawbacks in the present approach is a  systematic
overestimation of the energy positions of unoccupied spectral features
by about 25\%. We have attributed this to the use of the
linearized method employed for band structure calculations which is
expected to yield best results for the occupied states. 
The other limitation of the present description arises from 
a serious underestimation of the band gaps by the calculations. 
This necessitated the use of the single adjustable parameter
of the present approach which allowed us to shift the calculated 
spectra rigidly in order to match a peak in the experimental spectra.
In spite of this limitation, the agreement between the experiment
and calculation is convincing in view of the presence of
the single adjustable parameter, while the relative energy
positions, spectral intensities, widths and shapes are 
accurately described 
within an {\it ab-initio} method. The underestimation
of the band gap in LSDA approaches is well-known to arise 
from an average local treatment of the many-body potential
arising from electron-electron interactions and several
methods have been suggested to rectify  this \cite{Self energy}.
However, it 
appears that the effect of the Coulomb
interaction strength at the transition metal site on the
spectroscopic feature, is not as strong in these
trivalent compounds as in the divalent compounds of late
3$d$ transition elements arising from a lower value of $U/t$ 
in the trivalent case.
This is further supported by the 
near absence of any correlation induced satellite features
in any of the x-ray photoemission spectra of this series, while
intense satellite features are easily observed in such spectra of the
$M$O series. Moreover,
the spectral widths in the La$M$O$_3$ series  
appear to be dominated by band structure
effects rather than multiplet interaction effects,

\noindent {\bf Acknowledgment}

D.D.S thanks Dr. M. Methfessel, 
Dr. A. T. Paxton, and Dr. M. van Schiljgaarde for making
the LMTO-ASA band structure program available and Dr. S. Krishnamurthy
for initial help in setting up the LMTO-ASA program. 
We also thank Drs. N. Hamada, H. Sawada and K. Terakura for
several discussions and an earlier collaboration \cite{PRL} that 
preceded this work.
Dr. N. Shanthi
thanks the CSIR, Government of India for the Research
Associateship.

\begin{figure}
\caption{(a) Experimental valence band x-ray photoemission spectrum
(dots) of LaCrO$_3$ compared with the calculated spectrum (solid line).
The relative contributions from Cr $d$ states (dash) and O $p$ states
(dash-dot) to the calculated spectrum are also shown.
(b) The spin-polarized contributions of Cr $d$ states 
from one of the Cr sites to the calculated spectrum is shown here;
the solid line is the up-spin contribution, while dashes
represent the down-spin contribution.}
\label{Fig. 1}
\end{figure}

\begin{figure}
\caption{A wide scan BI spectrum of La$_{0.9}$Sr$_{0.1}$CrO$_3$ showing
the intense La 4$f$ peak at about 8.5 eV and weak intensity features
related to Cr $d$ states below 7 eV.}
\label{Fig. 2}
\end{figure}

\begin{figure}
\caption{(a) Experimental BI spectrum of
(dots) La$_{0.9}$Sr$_{0.1}$CrO$_3$ compared 
with the calculated spectrum (solid line) for LaCrO$_3$.
The relative contributions from Cr $d$ states (dash) and O $p$ states
(dash-dot) to the calculated spectrum are also shown.
(b) The spin-polarized contributions of Cr $d$ states from
one of the Cr sites to the calculated spectrum is shown here;
the solid line is the up-spin contribution, while dashes
represent the down-spin contribution.} 
\label{Fig. 3}
\end{figure}

\begin{figure}
\caption{ 
Experimental x-ray absorption spectrum (dots) of
La$_{0.9}$Sr$_{0.1}$CrO$_3$ at the oxygen 
$K$-edge compared 
with the calculated spectrum (solid line) for LaCrO$_3$. The
broadened spin-polarized Cr $d$ partial DOS are also 
shown: up-spin (dash) and down-spin (dash-dot). }
\label{fig4}
\end{figure}

\begin{figure}
\caption{(a) Experimental BI spectrum (dots) of 
LaMnO$_3$ compared with the calculated spectrum (solid-line).
The relative contributions from Mn $d$ states (dash) and O $p$ states
(dash-dot) to the calculated spectrum are also shown.
(b) The spin-polarized contributions of Mn $d$ states 
from one of the Mn sites to the calculated spectrum is shown here;
the solid line is the up-spin contribution, while dashes
represent the down-spin contribution.}
\label{Fig. 5}
\end{figure}

\begin{figure}
\caption{Experimental BI spectrum (dots) of LaFeO$_3$ compared 
with the calculated spectrum for LaFeO$_3$ (solid line).
The relative contributions from Fe $d$ states (dash) and O $p$ states
(dash-dot) to the calculated spectrum are also shown.
(b) The spin-polarized contributions of Fe $d$ states 
from one of the Fe sites to the calculated spectrum is shown here;
the solid line is the up-spin contribution, while dashes
represent the down-spin contribution.}
\label{Fig. 6}
\end{figure}

\begin{figure}
\caption{Experimental x-ray absorption spectrum (dots) of
LaFeO$_3$ at the oxygen 
$K$-edge compared 
with the calculated spectrum (solid line) for LaFeO$_3$. The
broadened spin-polarized Cr $d$ partial DOS are also 
shown: up-spin (dash-dot) and down-spin (dash).}
\label{Fig. 7}
\end{figure}

\begin{figure}
\caption{Experimental BI spectrum (dots) of LaCoO$_3$ compared 
with the calculated spectrum (solid line) for LaCoO$_3$.
The relative contributions from Co $d$ states (dash) and O $p$ states
(dash-dot) to the calculated spectrum are also shown.} 
\label{Fig. 8}
\end{figure}

\begin{figure}
\caption{Experimental BI spectrum (dots) of LaNiO$_3$ compared 
with the calculated spectrum (solid line) for LaNiO$_3$.
The relative contributions from Ni $d$ states (dash) and O $p$ states
(dash-dot) to the calculated spectrum are also shown.} 
\label{Fig. 9}
\end{figure}

\end{document}